\documentclass[12pt]{article}

\usepackage{amsmath, amsfonts, amssymb, amsthm, natbib}
\usepackage{graphicx,psfrag,epsf}
\usepackage{latexsym}
\usepackage{calc}
\usepackage[ragged, symbol]{footmisc}
\usepackage{grffile}
\usepackage{booktabs}
\usepackage{algorithmic}
\usepackage{algorithm}
\usepackage{caption}
\usepackage{comment}
\usepackage{color,url}
\usepackage{enumitem}
\usepackage{bbm}

\usepackage{bm}

\newcommand{\V}[1]{{\bm{\mathbf{\MakeLowercase{#1}}}}} % vector
\newcommand{\M}[1]{{\bm{\mathbf{\MakeUppercase{#1}}}}} % matrix
\usepackage[utf8]{inputenc} % allow utf-8 input
\usepackage[T1]{fontenc}    % use 8-bit T1 fonts
\usepackage{hyperref}       % hyperlinks
\usepackage{url}            % simple URL typesetting
\usepackage{booktabs}       % professional-quality tables
\usepackage{amsfonts}       % blackboard math symbols
\usepackage{nicefrac}       % compact symbols for 1/2, etc.
\usepackage{microtype}      % microtypography

\usepackage{amsmath}
\usepackage{multirow}
\usepackage{float}
\usepackage{txfonts}

\theoremstyle{plain}

\newtheorem{theorem}{Theorem}[section]

\begin{document}
%\bibliographystyle{natbib}

\def\spacingset#1{\renewcommand{\baselinestretch}%
{#1}\small\normalsize} \spacingset{1}

%%%%%%%%%%%%%%%%%%%%%%%%%%%%%%%%%%%%%%%%%%%%%%%%%%%%%%%%%%%%%%%%%%%%%%%%%%%%%%

%\if0\blind
%{
\title{Hierarchical Continuous Time Hidden Markov Model, with Application in Zero-Inflated Accelerometer Data}
 
 \author{Zekun Xu\thanks{
 Department of Statistics, North Carolina State University (Email: \textit{zxu13@ncsu.edu})}
 \and
 Eric B. Laber\thanks{
 Department of Statistics, North Carolina State University (Email: \textit{elaber@ncsu.edu})}
\and
Ana-Maria Staicu\thanks{
 Department of Statistics, North Carolina State University (Email: \textit{astaicu@ncsu.edu})}}
  \date{}
\maketitle

\bigskip
\begin{abstract}
Wearable devices including accelerometers are increasingly 
  being used to collect high-frequency human activity 
  data {\em in situ}.  There is tremendous potential
  to use such data to inform medical decision making 
  and public health policies.  However, modeling
  such data is challenging as they are high-dimensional,
  heterogeneous, and subject to informative 
  missingness, e.g., zero readings when the device
  is removed by the participant.  We propose a flexible
  and extensible continuous-time hidden Markov
  model to extract meaningful activity patterns from 
  human accelerometer data.   To facilitate estimation
  with massive data we derive an efficient learning
  algorithm that exploits the hierarchical structure
  of the parameters indexing the proposed model.  We
  also propose a bootstrap procedure for interval estimation.  
  The proposed methods are illustrated using data from the 
  2003 - 2004 and 2005 - 2006 National Health and Nutrition
  Examination Survey.
\end{abstract}
\newpage

\section{Introduction}

The development of the wearable technology has given rise to a variety
of sensing devices and modalities. Some of these devices,
e.g., Fitbit or Apple Watch, can be worn continuously 
and thereby produce huge volumes of high-frequency human activity data. 
Because these data present little burden on the wearer to collect and
provide rich information on the {\em in situ} behavior of the wearer, they
have tremendous potential to inform decision making in healthcare.  
Examples of remote sensing data in healthcare include  
elder care, remote monitoring of chronic disease,
and addition management \citep[][]{bartalesi2005wearable,marshall2007self,hansen2012accelerometer}. 
Accelerometers are among the most commonly used and most widely studied 
types of wearable devices, 
they have been used both in randomized clinical trials to
evaluate treatment effect on activity-related impairment
\citep[][]{napolitano2010accelerometer,kanai2018effect} and in
observational studies to characterize activity patterns in a
free-living environment 
\citep[][]{morris2006using,hansen2012accelerometer,xiao2014quantifying}.
However, despite rapidly growing interest and investment 
in wearable devices for the study of human activity data,
a general and extensible class of models for analysis of 
the resulting data is lacking.  

We propose a continuous-time hidden Markov model for the modeling
human accelerometer data that aligns with scientific 
(conceptual) models of human activity data; in the proposed 
model, latent states correspond to latent (unobserved) 
activities, e.g., resting, running, jumping, etc., that 
are shared across the population yet the accelerometer signatures 
within these activities are allowed to vary across subjects.  
Furthemore, differences across subgroups, e.g., defined by
sex, age, or the presence-absence of a comorbid condition, 
can be identified by aggregating individual-level effects
across these groups.  
This work is motivated in part by
the physical activity data set from the 2003 - 2004 National Health
and Nutrition Examination Survey (NHANES). In this study, human
activity patterns were measured at one-minute intervals for up to seven
days using the ActiGraph Model 7164 accelerometer
\citep[][]{troiano2008physical,metzger2008patterns,schmid2015associations}.
Activity for each minute was recorded as an integer-valued 
intensity-level commonly referred to as an activity count.    
In the study,  subjects were instructed to 
remove the device
during sleep or while washing (to keep it dry). 
Therefore, the observed data comprise high-frequency,
integer-valued activity counts for each subject with 
intervals of missing values
corresponding to when the device was removed. 

The goal of paper is to use the observed data to characterize 
activity patterns of each subject, subjects within pre-defined subgroups,
and the population as a whole. This is important because the estimated physical activity 
model can potentially serve the following three purposes.
On the subject level, the estimated activity model can be used both for the prediction of future activities
and the imputation of missing activity readings.
On the subgroup level, the estimated activity patterns provide useful insights into clustering people
based on their activity profiles.
On the population level, public policies can be designed based on the estimated activity model
so as to encourage everyday exercise and healthy life style.

Prior work on modeling activity counts has focused on aggregation
and other smoothing techniques.  
One common approach is to average the activity
counts over time for each subject and then compare the group means using
two-sample t-tests \citep[][]{troiano2008physical}, analysis of covariance
\citep[][]{hansen2012accelerometer}, or linear mixed effects models
\citep[][]{cradock2004youth}. However, in these approaches, 
averaging focuses on overall activity levels and may obscure trends
in activity type, activity duration, and transitions between activities.
Another approach is to use 
functional data analysis methods 
wherein the integer activity counts are first log-transformed to
fix the right-skewedness in the distribution and the transformed
activity modeled as a function of time of day and other
covariates \citep[][]{morris2006using,xiao2014quantifying,gruen2017use}. 
These approaches are best-suited for the identification of 
smooth, cyclical patterns in the data whereas the observed
accelerometer data are characterized by abrupt (i.e., non-smooth)
changes in activity levels.

Discrete-time hidden Markov models are another common
approach to the analysis of mobility data measured by wearable
devices \citep[][]{he2007real,nickel2011using,ronao2014human,witowski2014using}. In these
models, activity is partitioned into different latent
behavioral states and the observed activity count is dependent on the
unobserved latent activity. The latent states evolve according to a 
discrete-time Markov process and a primary goal
is the correct classification of the latent activity. To construct and
validate these models requires training data that are labeled
by latent activity.  However, the NHANES data, like many 
accelerometer studies, are not labeled by activity.  
Furthermore, our goal is to identify the dynamics of 
a patients evolution through these latent activities including
activity duration, activity intensity, and transitions between
activities.  
Discrete-time hidden Markov models have been used to
model latent health states and subsequently conduct inference for
activity patterns within each state
\citep[][]{scott2005hidden,altman2007mixed,shirley2010hidden}, but the time
scales in these applications are rather coarse (daily or weekly). By
contrast, the physical activity in the NHANES data is measured for
each minute; this results in a much larger data volume and the ability to
provide are more complete picture of activity dynamics.

A technical
limitation of the discrete-time approach, is that it assumes that 
the observations are
equally spaced in time. 
Continuous-time hidden Markov models
(CTHMM) have been used to analyze the irregularly-sampled temporal
measurements
\citep[][]{nodelman2012expectation,wang2014unsupervised,liu2015efficient}. 
The flexibility of the CTHMM comes at the expense of increased
computational cost, which makes it infeasible
for large datasets without modification. \cite{liu2015efficient}
developed an efficient learning algorithm
for parameter estimation in the CTHMMs. However, this
algorithm is only suitable for either the completely pooled or unpooled
cases wherein all subjects are assumed to be either completely
homogeneous so that they share the same parameters, or completely
heterogeneous so that all parameters are subject-specific. Moreover,
the algorithm cannot estimate the effects of subject
covariates and environmental factors on activity counts.  

We propose to model minute-by-minute accelerometer data using a
hierarchical continuous-time hidden Markov model (HCTHMM). This model
is aligned with scientific models of activity as the latent states represent
different types of unobserved physical activities. The 
continuous-time Markov process for the latent
states evolution avoids having to perform imputation for missing
yet allows for the possibility that the
temporal measurements are irregularly spaced. Furthermore, the proposed 
model can incorporate both baseline subject covariates and 
time-varying environmental factors. The proposed model
is hierarchical in that it is parameterized by: 
(i) subject-specific parameters
to account for variability between subjects; (ii) subgroup-specific
parameters parameters to account for similarity in activity
patterns within groups; (iii) population
parameters that are common across all subjects.  This specification
allows us to pool information on some parameters while
retaining between-group and between-subject variability. We
proposed an estimator of these parameters that 
is based on 
consensus optimization using the alternating direction method of
multipliers (ADMM). There is a vast literature on the convergence
properties of ADMM
\citep[][]{boyd2011distributed,shi2014linear,hong2017linear} which can be readily ported to the proposed algorithm. Finally, we use the 
nonparametric bootstrap \citep[][]{efron1992bootstrap} to estimate 
the sampling distributions of parameter estimators and to conduct
statistical inference.

\section{Model Framework}\label{modelsec}
We  assume that the observed data are of the form
$\{\M{W}_i,Y_i(\M{T}_i),\M{X}_i(\M{T}_i) \}_{i=1}^n$, which comprise
 $n$ independent copies of the trajectory
$\{\M{W}, Y(\M{T}), \M{X}(\M{T}) \}$, where: $\M{W}\in\mathbb{R}^p$
are baseline subject characteristics; 
$Y(\M{T})=\{Y(T_1),\ldots,Y(T_K)\}$ are the non-negative integer 
activity counts
at times $\M{T}=(T_1,\ldots,T_K) \in [0,1]^{K}$; and
$\mathbf{X}(\M{T})=\{\M{X}(T_1),\ldots,\M{X}(T_K)\}$ are
concurrent environmental factors such that $\M{X}(\cdot)
\in\mathbb{R}^q$.
Both $\M{T}$ and $K$ are treated as random variables as the number and
timing of observations vary across subjects.  We model the evolution
of the observed data using a hierarchical continuous-time hidden
Markov model (HCTHMM), which we will develop over the remainder of this
section.

Let $S_i(t)\in\{1,\ldots,M\}$ denote the unobserved latent state for
subject $i=1,\ldots,n$ at time $t\in[0,1]$. The latent state 
evolves according to a Markov process indexed by: (i) an initial state
distribution $\pi_i(m) \triangleq P\{S_i(0)=m)\}$ for $m=1,\ldots,M$
such that $\sum_{m=1}^M\pi_i(m)=1$; (ii) a transition rate matrix
$\M{Q}_i=\{q_i(m,\ell)\}_{m,\ell=1,\ldots,M}$ such that
$q_i(m,m)=-\sum_{\ell\ne m}q_i(m,\ell)$. The transition rate matrix, 
also known as the infinitesimal generator matrix, describes the
rate of movements between states in a continuous-time Markov chain
\citep[][]{pyke1961markova,pyke1961markovb,albert1962estimating}; 
the transition probabilities are derived from the
transition rates through a matrix exponential operation such that for
$k=1,\ldots,K-1$ and $t > u$, 
$P_i^{t-u}(m,\ell)\triangleq P\{S_i(t)=\ell |
S_i(u)=m\}=\{e^{(t-u) \M{Q}_i}\}_{m,\ell}$.

We assume that the conditional distribution of the activity counts is
homogeneous in time given the current latent state and environmental
factors (to streamline notation, we include baseline characteristics
in the time-varying environmental factors).  For $i=1,\ldots,n$ and
$m=1,\ldots, M$ define
$g_{i,m}(y; \mathbf{x}) \triangleq
P\{Y_i(t)=y|S_i(t)=m,\mathbf{X}_i(t)=\mathbf{x}\}$.
Because longitudinal activity count data are zero-inflated,
we set $g_{i,1}(y; \mathbf{x})$ to be the probability mass function for a
zero-inflated Poisson distribution with structural zero proportion
$\delta_i$ and mean $\lambda_{i,1}$ for state 1 such that
\begin{equation*}
g_{i,1}(y; \mathbf{x}) = \delta_{i}(\mathbf{x})1_{y=0} + \left\lbrace
1-\delta_{i}(\mathbf{x})
\right\rbrace 
\frac{\left\lbrace \lambda_{i,1}(\mathbf{x})\right\rbrace^{y}\exp\left\lbrace
\lambda_{i,1}(\mathbf{x})
\right\rbrace}{
y!
}.
\end{equation*}
For $m=2,\ldots, M$ we model the activity counts using a 
Poisson
regression model so that
\begin{equation*}
g_{i,m}(y; \mathbf{x})=\frac{\left\lbrace \lambda_{i,m}(\mathbf{x})\right\rbrace^{y}\exp\left\lbrace
\lambda_{i,m}(\mathbf{x})
\right\rbrace}{
y!
}.
\end{equation*}
 For each subject $i$, and latent
state $m$,
we assume that functions $\delta_i(\mathbf{x})$ and $\lambda_{i,m}(\mathbf{x})$are of the form
\begin{eqnarray*}
\log\left\lbrace
\frac{\delta_i(\mathbf{x})}{1-\delta_i(\mathbf{x})} 
\right\rbrace &=& b_{i,0,0} + \mathbf{x}'\mathbf{b}_{i,0,1}, \\
\log\left\lbrace
\lambda_{i,m}(\mathbf{x})
\right\rbrace &=& b_{i,m,0}+\mathbf{x}'\mathbf{b}_{i,m,1},
\end{eqnarray*}
where $b_{i,0,0},\ldots,b_{i,M,0}$ and
$\mathbf{b}_{i,0,1},\ldots,\mathbf{b}_{i,M,1}$ are unknown coefficients.

In the foregoing model description, all parameters are
subject-specific so that each subject's trajectory can be modeled
separately. However, in the HCTHMM, some of the parameters are shared
among pre-defined subgroups of the subjects.  
We assume that subjects are partitioned into
$J$ such subgroups based on their baseline characteristics $\mathbf{W}$. For
example, these groups might be determined by age and sex. 
Subjects within the same group are
though to behave more similarly to each other than across groups. Let
$G_i\in\{1,\ldots,J\}$ be the subgroup to which subject $i$ belongs, and 
let $n_j$ denote the number of subjects in group $j=1,\ldots, J$.

The HCTHMM is a flexible multilevel model in that it allows
for three levels of 
of parameters: (i) subject-specific, (ii) subgroup-specific, (iii)
population-level. For example, one might let the intercepts in the generalized
linear models for state-dependent parameters be subject-specific
to account for the between-subject variability; let the
initial state probabilities and the transition rate parameters 
depend on 
group-membership, i.e. $\V{\pi}_{i_1}=\V{\pi}_{i_2}$ and
$\M{Q}_{i_1}=\M{Q}_{i_2}$ for all $i_1,i_2$ such that
$G_{i_1}=G_{i_2}$; and let the slope parameters in the generalized
linear models for state-dependent parameters be common across all
subjects.

If all the observed time points are equally spaced and all the
parameters are subject-specific, the HCTHMM reduces to the
subject-specific zero-inflated Poisson hidden Markov model.  If there
are no covariates and all parameters are common for all subjects, the
HCTHMM reduces to a zero-inflated variant of the continuous-time
hidden Markov model \citep[][]{liu2015efficient}. The extension from the
previous models to the HCTHMM better matches the scientific goals
associated with analyzing the NHANES data but also requires new
methods for estimation.  Because of the hierarchical structure in the
parameters, joint parameter estimation is no longer embarrassingly
parallelizable as it would be in the case of its completely pooled or
unpooled counterparts.  

\section{Parameter Estimation}

\subsection{Forward-Backward Algorithm}\label{likelihood}

For subject $i=1,\ldots,n$, let $\V{a}_i\in\mathbb{R}^{M-1}$ be the $M-1$
free parameters in the initial probabilities $\V{\pi}_i$, and
let $\V{c}_i\in\mathbb{R}^{M(M-1)}$ be the $M(M-1)$ free 
parameters in the transition matrix $\M{Q}_i$. To simplify
notation, write
$\V{b}_{i,0}\triangleq [b_{i,0,0},\ldots,b_{i,M,0}]\in\mathbb{R}^{M+1}$
and
$\V{b}_{i,1}\triangleq
[\V{b}_{i,0,1}',\ldots,\V{b}_{i,M,1}']\in\mathbb{R}^{q(M+1)}$
to denote the parameters indexing the generalized linear models for
the activity counts in each state. Define the entire vector 
of parameters for subject $i$ to be
$\V{\theta}_i\triangleq
[\V{a}_i',\V{c}_i',\V{b}_{i,0}',\V{b}_{i,1}']\in\mathbb{R}^{(M+1)(M+q)}$.
The likelihood function for subject $i$ is computed using the
forward-backward algorithm 
\citep[][]{rabiner1989tutorial} as follows.
For subject $i=1,\ldots,n$, define the forward variables 
for $k = 1,\ldots,K-1$, and $m=1,\ldots,M$,
\begin{align*}
\alpha_i^{T_k}(m;\V{\theta}_i) \triangleq & 
P_{\V{\theta}_i}
\bigg\lbrace 
Y_i(T_1),\ldots,Y_i(T_k), S_i(T_k)=m \big| \pmb{X}_i(T_1)=\V{x}_{T_1},\nonumber\\
&\ldots, \pmb{X}_i(T_k)=\V{x}_{T_k}\bigg\rbrace.
\end{align*}
The initialization and recursion formulas are defined as
\begin{align*}
\alpha_i^{T_1}(m;\V{\theta}_i) = &\pi_i(m;\V{\theta}_i) g_{i,m}\{Y_i(T_1);\V{x}_{T_1},\V{\theta}_i\},\\
\alpha_i^{T_{k+1}}(m;\V{\theta}_i)  = &\left[ \sum_{\ell=1}^M\alpha_i^{T_k}(\ell;\V{\theta}_i) \{e^{(T_{k+1}-T_{k}) \M{Q}_i}\}_{\ell, m} \right]g_{i,m}\{Y_i(T_{k+1}); \nonumber \\
& \V{x}_{T_{k+1}},\V{\theta}_i\},
\end{align*}
where $m=1,\ldots,M$ and $k=1,\ldots,K-1$. The 
negative log-likelihood for $\V{\theta}_i$ is therefore
$f_i(\V{\theta}_i) = -log\left\lbrace
\sum_{m=1}^{M}\alpha_i^{T_K}(m;\V{\theta}_{i})
\right\rbrace$. Define the joint likelihood for 
$\V{\theta} = (\V{\theta}_{1},\ldots, \V{\theta}_{n})$ to be
$f(\V{\theta}) = \sum_{i=1}^{n}f_i(\V{\theta}_{i})$.

To compute the conditional state probabilities in the HCTHMM, we need
to generate a set of auxiliary backward variables analogous to the
forward variables defined previously. For subject $i=1,\ldots,n$,
define the backward variables for $k = 1,\ldots,K$, and $m=1,\ldots,M$,
\begin{align*}
\beta_i^{T_k}(m;\V{\theta}_i) \triangleq & 
P_{\V{\theta}_i}\bigg\lbrace Y_i(T_{k+1}),\ldots,Y_i(T_K) \big| S_i(T_k)=m, 
\nonumber \\
&\pmb{X}_i(T_{k+1})=\V{x}_{T_{k+1}},\ldots, \pmb{X}_i(T_K)=\V{x}_{T_K}
\bigg\rbrace.
\end{align*}
The initialization and recursion formulas are 
\begin{align*}
\beta_i^{T_K}(m;\V{\theta}_i) = &1,\\
\beta_i^{T_k}(m;\V{\theta}_i) =& 
\sum_{\ell=1}^M \{e^{(T_{k+1}-T_k) \M{Q}_i}\}_{m,\ell}\, 
g_{i,\ell}\{Y_i(T_{k+1});\V{x}_{T_{k+1}},\V{\theta}_i\} \nonumber \\
&\beta_i^{T_{k+1}}(\ell;\V{\theta}_i),
\end{align*}
where $m=1,\ldots,M$.
The probability of state $m$ for subject $i$ at time $t$ is 
\begin{align*}
\gamma_i^t(m;\V{\theta}_i) &\triangleq 
P_{\V{\theta}_i}\{S_i(t)=m|Y_i(T_1),\ldots,Y_i(T_K) \} \nonumber \\
&=\frac{\alpha_i^t(m;\V{\theta}_i)\beta_i^t(m;\V{\theta}_i)}{\sum_{m=1}^M\alpha_i^t(m;\V{\theta}_i)\beta_i^t(m;\V{\theta}_i)},
\end{align*}
where $t=T_1,\ldots,T_k$, $m=1,\ldots,M$, and $i=1,\ldots,n$.
The mean probability of state $m$ among subjects in group $j$ is thus  
\begin{align*}
\phi_j(m;\V{\theta}) &= \frac{1}{n_j}\sum_{\{i:G_i=j\}}\eta_i(m;\V{\theta}_i),\;j=1,\ldots,J, \\
\textrm{where}\;\eta_i(m;\V{\theta}_i)&=\frac{1}{K} \sum_{k=1}^{K} \gamma_i^{T_k}(m;\V{\theta}_i),\;m=1,\ldots,M. \nonumber
\end{align*}
The mean state probabilities $\phi_j(m;\V{\theta})$ can be interpreted as the mean proportion of time spent in latent state $m$ for subjects in group $j$, whereas $\eta_i(m;\V{\theta}_i)$ represents the mean proportion of time spent in state $m$ for subject $i$.

\subsection{Consensus Optimization}

If all sets of parameters are subject-specific, then the maximum likelihood
estimates for the parameters can be obtained by minimizing
$f(\V{\theta})$ using the gradient-based methods which can
be parallelized across subjects. However,
in the general setting where parameters are
shared across subgroups of subjects, such paralellization 
is no longer possible. 
Instead, we use
the consensus optimization approach to obtain the
maximum likelihood estimates in the HCTHMM, which is performed via the
alternating direction method of multipliers (ADMM)
\citep[][]{boyd2011distributed}. We use the Bayesian Information Criterion
(BIC) to select the number of latent states $M$.

Let $\M{D}$ denote a contrast matrix such that
$\M{D}\V{\theta} = 0$ corresponds to equality of 
subgroup-specific parameters within each subgroup and equality of 
all population-level parameters across all subjects.   
The maximum likelihood
estimator solves  
\begin{equation*}
\underset{\V{\theta}}{\min}f(\V{\theta})\;\;s.t.\;\; \M{D}\V{\theta}=0.
\end{equation*}
For the purpose of illustration, 
suppose that: (i) the intercepts in the
generalized linear models for state-dependent parameters are
subject-specific; (ii) the initial state probabilities and the
transition rate parameters are subgroup-specific; (iii) the slope
parameters in the generalized linear models for state-dependent
parameters are common across all subjects. Then $\M{D}\V{\theta}=0$
is the same as restricting (i) $\V{a}_{i_1} = \V{a}_{i_2}$ for all
$G_{i_1}=G_{i_2}$; (ii) $\V{c}_{i_1} = \V{c}_{i_2}$ for all
$G_{i_1}=G_{i_2}$; (iii) $\V{b}_{i,1} = \V{b}_1$.

In our illustrative example, the maximum likelihood estimator solves 
\begin{align*}
&\underset{\V{\theta},\V{z}}{\min}f(\V{\theta})\\
&s.t.\;\M{A}_i\V{\theta}_i = \M{B}_i\V{z},\;i=1,\ldots,n, \nonumber
\end{align*}
where $\V{z}$ represents the set of all subgroup-specific and common
parameters in $\V{\theta}$ so the linear constraint
$\M{A}_i\V{\theta}_i = \M{B}_i\V{z}$ is equivalent to
$\M{D}\V{\theta}=0$. 
%A concrete example of $\M{A}_i$ and $\M{B}_i$ is
%given in the Supplementary Materials.
The corresponding augmented Lagrangian is
\begin{align*}
L_\rho(\V{\theta},\V{z},\V{\xi})&=f(\V{\theta})+\V{\xi}^T(\M{A}\V{\theta}-\M{B}\V{z})+\frac{\rho}{2}\|\M{A}\V{\theta}-\M{B}\V{z}\|_2^2\\
&=\sum_{i=1}^n\{f_i(\V{\theta}_i)+\V{\xi}_i^T(\M{A}_i\V{\theta}_i-\M{B}_i\V{z})+\frac{\rho}{2}\|\M{A}_i\V{\theta}_i-\M{B}_i\V{z}\|_2^2    \} , \nonumber 
\end{align*}

\begin{align*}
\textrm{where }&\V{\xi}=[\V{\xi}_1',\ldots,\V{\xi}_n'],\; \nonumber \\
                &  \M{A}=\begin{bmatrix}\M{A}_1 & 0 & \cdots & 0 \\
                                                                                               0 & \M{A}_2 & \cdots & 0 \\
                                                                                               \vdots & &\ddots & 0\\
                                                                                               0 & 0 & \cdots &\M{A}_n \end{bmatrix},
                                                            \M{B}=\begin{bmatrix}\M{B}_1  \\
                                                                                              \M{B}_2  \\
                                                                                               \vdots \\
                                                                                               \M{B}_n \end{bmatrix}.
\end{align*}
Here $\V{\xi}$ are the Lagrange multipliers and $\rho$ is a pre-specified
positive penalty parameter. Let
$\tilde{\V{\theta}}^{(v)}_n,\tilde{\V{z}}^{(v)}_n,\tilde{\V{\xi}}^{(v)}_n$
be the $v^{th}$ iterates of $\V{\theta},\V{z},\V{\xi}$. Then, at the
iteration $v+1$, the ADMM algorithm updates are
\begin{align*}
&\V{\theta}-\textrm{update: } \nabla f(\tilde{\V{\theta}}^{(v+1)}_n)+\M{A}^T\tilde{\V{\xi}}^{(v)}_n+\rho\M{A}^T(\M{A}\tilde{\V{\theta}}^{(v+1)}_n-\M{B}\tilde{\V{z}}^{(v+1)}_n)=0,\\
&\V{z}-\textrm{update: } \M{B}^T\tilde{\V{\xi}}^{(v)}_n+\rho\M{B}^T(\M{A}\tilde{\V{\theta}}^{(v+1)}_n-\M{B}\tilde{\V{z}}^{(v)}_n)=0,\\
&\V{\xi}-\textrm{update: } \tilde{\V{\xi}}^{(v+1)}_n-\tilde{\V{\xi}}^{(v)}_n-\rho(\M{A}\tilde{\V{\theta}}^{(v+1)}_n-\M{B}\tilde{\V{z}}^{(v+1)}_n)=0,
\end{align*}
where the most computationally expensive $\V{\theta}$-update 
can be programmed in parallel across each $i=1,\ldots,n$ as
\begin{equation*}
\V{\theta}_i-\textrm{update: } \nabla f_i(\tilde{\V{\theta}}_{i,n}^{(v+1)})+\M{A}_i^T\tilde{\V{\xi}}_{i,n}^{(v)}+\rho\M{A}_i^T(\M{A}_i\tilde{\V{\theta}}_i^{(v+1)}-\M{B}_i\tilde{\V{z}}^{(v)}_n)=0.
\end{equation*}
The gradients $\nabla f_i(\cdot)$ for subject $i$'s HMM parameters can
be computed using Fisher's identity \citep[][]{cappe2005inference} based on
the efficient EM algorithm proposed in \cite{liu2015efficient}. The
details are included in the Supplementary Materials.  The use of
gradients is needed in our model both due to the ADMM update and the
covariate structure. Even when there are no covariates and all
parameters are shared across subjects, the gradient method is still
faster than the EM algorithm because 
M-step is expensive in the zero-inflated Poisson
distribution.

\subsection{Theoretical Properties}

Define $\hat{\V{\theta}}_n$ to be the maximum likelihood
estimator for $\V{\theta}$ and let $\V{\theta}^\star$ denote
its population-level analog. The following
are the sufficient conditions to ensure: (i) almost sure
convergence of $\hat{\V{\theta}}_n$ to $\V{\theta}^\star$ as
$n\to\infty$, and (ii)  numerical convergence of
$\tilde{\V{\theta}}_n^{(v)}$ to $\hat{\V{\theta}}_n$ as $v\to\infty$.

\bigskip
\noindent
\begin{enumerate}
\item [(A0)] The true parameter vector $\V{\theta}^\star$ for the
  unconstrained optimization problem
  $\underset{\V{\theta}}{\min}f(\V{\theta})$ is an interior point of
  $\mathbf{\Theta}$, where $\mathbf{\Theta}$ is a compact subset of
  $\mathbb{R}^{\dim\,\V{\theta}}$.
\item [(A1)] The constraint set
  $\mathcal{C}\triangleq\{\V{\theta}\in\mathbf{\Theta};
  \M{D}\V{\theta}=0 \}$
  is nonempty and for some $r\in\mathbb{R}$, the set
  $\{\V{\theta}\in C; f(\V{\theta})\leq r\}$ is nonempty and
  compact.%feasible.
\item [(A2)] The observed time process $(T_k: k\in\mathbb{N})$ is
  independent of the generative hidden Markov process: the likelihood
  for the observed times do not share parameters with $\V{\theta}$.
%still a discrete time process (view as inhomogeneous)
\item [(A3)] There exist positive real numbers $0<\kappa^-\leq\kappa^+<1$ such that for all subjects $i=1,\ldots,n$,
$\kappa^- \leq P_i^{T_{k+1}-T_k}(m,\ell) \leq \kappa^+$  
for $k=1,\ldots, K-1$ and $m,\ell=1,\ldots, M$, almost surely and 
$g_{i,m}(y;\V{x})>0$  for all $y\in\textrm{ supp }Y$ for some 
$m=1,\ldots,M$.
\item [(A4)] For each $\V{\theta}\in\mathbf{\Theta}$, the transition
  kernel indexed by $\V{\theta}$ is Harris recurrent and
  aperiodic. The transition kernel is continuous in $\pmb{\theta}$ in
  an open neighborhood of $\pmb{\theta}^\star$.
\item [(A5)] The hidden Markov model is identifiable up to label switching of the latent states.
\item [(A6)] The negative log likelihood function $f(\V{\theta})$ is
  twice differentiable with respect to $\V{\theta}$ with bounded,
  continuous derivatives. Denote by $e_{min}(\V{\theta})$ 
and $e_{max}(\V{\theta})$ to be the
  smallest and largest eigenvalues of $\nabla^2f(\V{\theta})$ then 
  there exists positive real numbers
  $0<\varrho_-\leq\varrho_+<\infty$ such that $e_{min}(\V{\theta})
  \geq\varrho_->0$
  and $e_{max}(\V{\theta})\leq\varrho_+<\infty$ for all
  $\V{\theta}\in\V{\Theta}$. 
\end{enumerate}
%Define the Hessian matrix$$\mathcal{H}(\V{\theta}^\star) \triangleq\frac{\partial^2 f(\V{\theta})}{\partial \V{\theta} \partial \mathbf{\theta}'}|_{\V{\theta}=\V{\theta}^\star}$$ is nonsingular at the true parameter value $\V{\theta}^\star$.
Assumption (A0)-(A2) are mild regularity conditions whereas
(A3) - (A5) are standard in 
latent state space models; together they ensures that the model is
well-defined. Assumption (A4) is a avoids
non-standard asymptotic behavior associated with non-smooth
functionals.  Assumption (A6) can be used to show the Lipschitz
continuity in the gradient and strong local convexity
which are used to establish the numerical convergence in the
ADMM algorithm \cite{shi2014linear}. 
\begin{theorem}
Under assumptions (A0) - (A6), as $T_i\to\infty$ for $i=1,\ldots,n$,
\begin{enumerate}
\item[(i)] $\hat{\V{\theta}}_n$ converges to $\V{\theta}^\star$ almost surely as $n\to\infty$, 
\item[(ii)] $\tilde{\V{\theta}}_n^{(v)}$ converges numerically to $\hat{\V{\theta}}_n$ in an open neighborhood of $\V{\theta}^\star$ as $v\to\infty$.
\end{enumerate}
\end{theorem}

The first part of Theorem 1 states the almost sure convergence
of the constrained maximum likelihood estimator $\hat{\V{\theta}}_n$
to the true parameter value $\V{\theta}^\star$. This can be shown
using the uniform convergence results of the log likelihood
\citep[][]{douc2004asymptotic} for each subject-specific hidden Markov
model, along with the feasibility assumption (A1) and identifiability
assumption (A5). The second part of Theorem states the numerical
convergence of the ADMM algorithm. This is anticipated by 
\cite{boyd2011distributed}
which identifies general conditions for the numerical convergence of the
residual, the dual variable, and the objective
function. \cite{shi2014linear} extended
the convergence to the primal variable by adding the Lipschitz
continuity and strong convexity assumptions. The details for those
assumptions, as well as the proof for Theorem 1, are included in the
Supplementary Materials. 

% Because the inference of HCTHMM is based on the mean proportion of
% time spent in each latent state among subjects in each group, we are
% interested in the asymptotic properties of the mean state
% probabilities. 
For $i=1,\ldots,n$, define the estimator for the mean
proportion of time in state $m$ in group $k$ as
$\hat{\phi}_{j,n}(m), j=1,\ldots,J, m=1,\ldots,M$, and the estimator for the
mean proportion of time in state $m$ as $\hat{\eta}_{i,k_i}$ where
$k_i$ is the number of observed time points for subject $i$.  
%Denote
%$\overset{a.s.}{\longrightarrow}$ to be almost sure convergence and
%$\overset{d}{\longrightarrow}$ to be convergence in distribution.
The following result characterizes the limiting behavior of the
estimated time in each state.  
\begin{theorem}
Under (A0) - (A6), as $k_i\to\infty$ for $i=1,\ldots,n$, $n_j\to\infty$ for $j=1,\ldots,J$,
\begin{itemize}
\item[(i)] $\hat{\phi}_{j,n}(m;\hat{\V{\theta}}_n)$
converges to 
$\mu_j(m;\V{\theta}^\star)$ almost surely,
\item[(ii)] $\frac{\hat{\phi}_{j,n}(m;\hat{\V{\theta}}_n) - \mu_j(m;\hat{\V{\theta}}_n)}{\sqrt{\sigma_j^2(m;\hat{\V{\theta}}_n)/n_j}}$ converges 
in distribution to a standard normal random variable,
 where 
$\mu_j(m;\hat{\V{\theta}}_n)=E[\hat{\eta}_{i,k_i}(m;\hat{\V{\theta}}_n)],\;\sigma_j^2(m;\hat{\V{\theta}}_n)=Var[\hat{\eta}_{i,k_i}(m;\hat{\V{\theta}}_n)]$ for all $i$ such that $G_i=j.$ 
\end{itemize} 
\end{theorem}
%\|\gamma_i^t  - \bar{\gamma_i^t} \|_{TV} \leq \| \pi_k - \bar{\pi}_k \|_{TV} \leq \frac{\xi^+}{\xi^-}\left( 1 - \frac{\xi^-}{\xi^+} \right)^t \| \pi_0 - \bar{\pi_0} \|_{TV}
%difference in smoother no greater than difference in filter (projection)%
\noindent
A proof of the preceding result is given in the Supplementary
Materials, which follows from the almost sure convergence of a bounded
continuous function and the central limit
theorem. In principle, $\mu_j(m;\hat{\V{\theta}}_n)$ can be obtained from the
limiting distribution of a stationary continuous-time Markov chain,
which is determined by the transition as a function of
$\hat{\V{\theta}}_n$. However, it is generally not easy to compute 
the standard error analytically for the estimated mean state
probabilities. Instead, we use a stratified nonparametric bootstrap 
\citep[][]{efron1992bootstrap} in which we resample subjects with replacement 
from each subgroup.

\section{Simulation experiments}

We study the finite sample performance of the proposed estimator for
the state probabilities using a suite of simulation experiments.  We
simulate minute-by-minute activity counts of length
$T\sim \textrm{Uniform}(500, 2500)$ for $n=20$ and $n=200$ subjects,
where half of the subjects are male (Group 1) and the other half
female (Group 2).  The intervals between consecutive time points are
independently drawn from $\{1,2,\ldots,10\}$ with equal probabilities.
For each subject, we assume 2/7 of the observations are from weekends
and 5/7 of the observations are from weekdays.

The activity counts are generated using a three state continuous-time
zero-inflated Poisson hidden Markov model. We assume that during the
weekend the log mean activity decreases by 10\%, 20\%, 30\% in states
1, 2, 3 respectively, while the log odds of zero in state 1 increases
by 10\%, so that

\begin{eqnarray*}
\log\left\lbrace
\frac{\delta_i(\mathbf{x})}{1-\delta_i(\mathbf{x})} 
\right\rbrace &=& b_{i,0,0} + 0.1\times\mathbf{I}\{\textrm{Weekend}\}_i^{t}, \\
\log\left\lbrace
\lambda_{i,1}(\mathbf{x})
\right\rbrace &=& b_{i,1,0}- 0.1\times\mathbf{I}\{\textrm{Weekend}\}_i^{t},\\
\log\left\lbrace
\lambda_{i,2}(\mathbf{x})
\right\rbrace &=& b_{i,2,0} - 0.2\times\mathbf{I}\{\textrm{Weekend}\}_i^{t},\\
\log\left\lbrace
\lambda_{i,3}(\mathbf{x})
\right\rbrace &=& b_{i,3,0}-0.3\times\mathbf{I}\{\textrm{Weekend}\}_i^{t},
\end{eqnarray*}
where $b_{i,0,0}\overset{iid}{\sim}N(-1, 0.1^2),b_{i,1,0}\overset{iid}{\sim}N\left\lbrace \log(50), 0.1^2\right\rbrace,b_{i,2,0}\overset{iid}{\sim}N\left\lbrace \log(300),0.1^2\right\rbrace,b_{i,3,0}\overset{iid}{\sim}N\left\lbrace \log(700),0.1^2\right\rbrace$ are subject-specific intercepts; the weekend effect is assumed to be common across all subjects.

The initial probabilities for male are $(U_1,U_2,1-U_1-U_2)$, where $U_1,U_2\overset{iid}{\sim}$Uniform$(0.2,0.4)$; for female, the initial probabilities are $(U_3,U_4,1-U_3-U_4)$, where $U_3\overset{iid}{\sim}$Uniform$(0.6,0.8)$, $U_4\overset{iid}{\sim}$Uniform$(0.1,0.2)$. The transition rate matrix for male is  
$$\begin{bmatrix} -U_5-U_6 & U_5 & U_6 \\ U_7 & -U_7-U_8 & U_8 \\ U_9 & U_{10} & -U_9-U_{10} \end{bmatrix},$$
where $U_5,\ldots,U_{10}\overset{iid}{\sim}\mathrm{Uniform}(0.05,0.15)$; 
for female, the transition rate matrix is 
\noindent
$$\begin{bmatrix} -U_{11}-U_{12} & U_{11} & U_{12} \\ U_{13} & -U_{13}-U_{14} & U_{14} \\ U_{15} & U_{16} & -U_{15}-U_{16} \end{bmatrix},$$ 
where $U_{11},U_{12}\overset{iid}{\sim}\mathrm{Uniform}(0.05,0.1)$, 
$U_{13},U_{15}\overset{iid}{\sim}\mathrm{Uniform}(0.3,0.4)$, and  $U_{14},U_{16}\overset{iid}{\sim}\mathrm{Uniform}(0.1,0.2)$.

\begin{table}[H]
\small
\centering
\begin{tabular}{lllll}
\hline
Parameter & n=20 & n=200 \\ 
& Bias (s.e.) & Bias (s.e.) \\\hline
Population parameters: & & \\
Slope for State 1 zero odds  & .0021 (.0422) & .0032 (.0134) \\
Slope for State 1 Poisson mean & .0001 (.0034)& .0001 (.0011)\\
Slope for State 2 Poisson mean & .0016 (.0053) & .0019 (.0018)\\
Slope for State 3 Poisson mean & .0036 (.0120)& .0046 (.0044) \\
 \\
Subgroup-specific parameters: & & \\
Initial probabilities (Male)  & .0169 (.0022)& .0165 (.0008)\\
Initial probabilities (Female)  & .0152 (.0069) & .0284 (.0038)\\
Transition rates (Male) & .0038 (.0029) & .0011 (.0006)\\
Transition rates (Female) & .0099 (.0089) & .0016 (.0012) \\
 \hline      
\end{tabular}
\caption{The bias and standard error for the estimated population and subgroup-specific parameters in HCTHMM. The metric is Euclidean norm for population parameters, and Frobenius norm for subgroup-specific parameter vectors.}
\label{sim1}
\end{table}

\begin{figure}[H]
\centering
\includegraphics[scale=0.4]{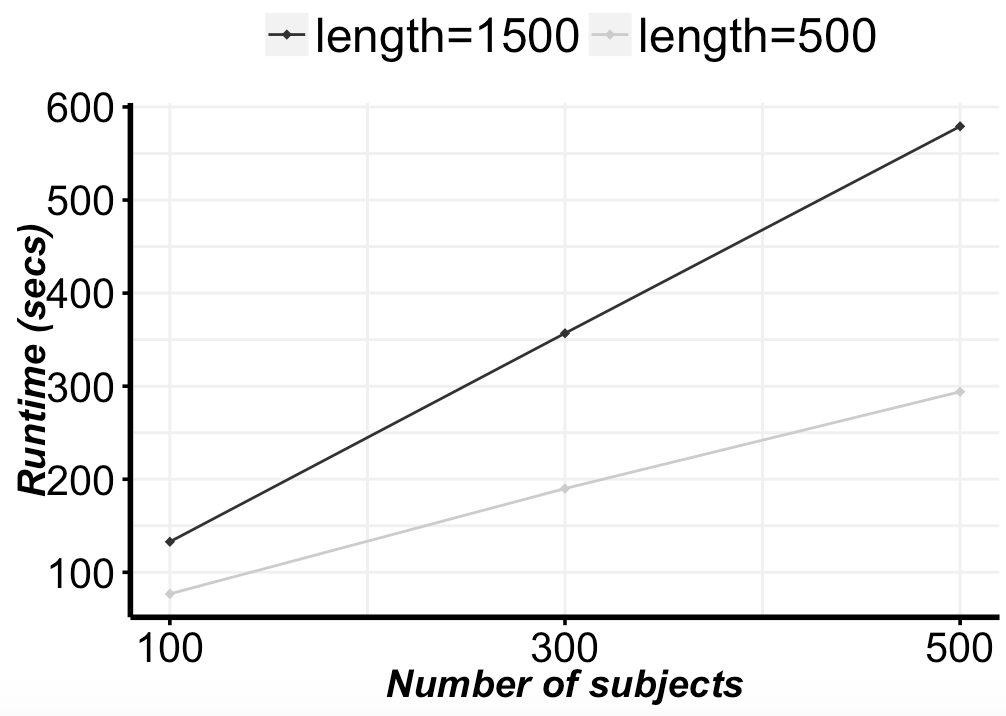}
\caption{\label{fig:your-figure} Average runtime (seconds) for each ADMM iteration in different configurations (number of subjects, length of each series). }
\label{timeplot}
\end{figure}

\begin{table}[H]
\small
\centering
\begin{tabular}{lllll}
\hline
\multicolumn{1}{l}{State} & \multicolumn{2}{c}{Subject-specific HMM} & \multicolumn{2}{c}{HCTHMM} \\
\cline{2-5}
\multicolumn{1}{l}{}  & \multicolumn{1}{p{1cm}}{$Male$} & \multicolumn{1}{p{1cm}}{$Female$} & \multicolumn{1}{p{1cm}}{$Male$} & \multicolumn{1}{p{1cm}}{$Female$}\\ \hline
95\% C.I.: & & &\\
1 &  .942 & .942  & .954  & .952 \\
2 &  .944 & .938 & .960  & .950 \\ 
3 &  .946 & .936 & .944  & .962 \\
\\
99\% C.I. : & & & \\
1 & .982 & .988 & .986   & .994\\
2 & .986 & .984  & .986   & .988\\
3 & .978 & .980  & .994  & .992\\
\hline      
\end{tabular}
\caption{Comparisons on the mean coverage probability for the 95\% and 99\% bootstrap confidence intervals 
for the mean proportion of time in each latent state between subject-specific HMM and HCTHMM (n = 200).}
\label{sim2}
\end{table}

Table \ref{sim1} shows the bias and standard error
of the estimators for different hierarchies of parameters in the
HCTHMM via 500 simulations. In both cases, the biases are small due to the fact that the
length of each individual series is large. As the sample size
increases, the standard errors become smaller which is
expected. Figure \ref{timeplot} shows the average runtime (seconds)
for each ADMM iteration scales linearly with the number of
subjects. It generally takes some 30 to 100 iterations for the
algorithm to converge.
Table \ref{sim2} compares the mean coverage probability of the 95\% and 99\%
bootstrap confidence intervals for the mean proportion of time
in each latent state bewteen a baseline subject-specific HMM and the proposed HCTHMM
when the sample size is 200. As we can see,
the baseline subject-specific HMM suffers undercoverage 
(coverage probability smaller than nominal level) in some of the latent states,
while the proposed HCTHMM recovers the nominal level well in
both the 95\% and 99\% cases.
  
\section{Application}\label{datasec}
 
The motivating application is a human
physical activity data set from the 2003 - 2004 National Health and
Nutrition Examination Survey (NHANES), which is publicly available at
the National Center for Disease Control (CDC) website
\url{https://wwwn.cdc.gov/Nchs/Nhanes/2003-2004/PAXRAW_C.htm}. 
There are 7,176 participants in the study, and for each participant we have 
minute-by-minute activity counts for up to seven days.
As the subjects were supposed to remove the accelerometer when washing,
there are prolonged intervals during the day when
accelerometer readings are zeros. We further impose the following two 
inclusion / exclusion criteria, 
\begin{itemize}
\item Subjects whose age is between 20 and 60 are included.
\item Subjects with very few measurements are excluded.
\end{itemize}
The first criterion specifies the scope of inference. The second criterion exclude subjects with very few non-missing data available ($<$ 500 minutes out of 7 days). There are 2,467 subjects who satisfy both conditions, which constitute more than 95\% of those whose age is between 20 and 60. Further, we split those subjects by their baseline characteristics (gender, age) into 4 subgroups. Subgroup 1 consists of 608 male subjects with age from 20 to 40; subgroup 2 consists of 557 male subjects with age from 40 to 60; subgroup 3 consists of 712 female subjects with age from 20 to 40; and subroup 4 consists of 590 female subjects with age from 40 to 60.

\begin{table}[H]
\small
\centering
\begin{tabular}{lc}
\hline
Literature & Definition of missing\\ \hline
\cite{cradock2004youth} & 30 minutes \\
\cite{catellier2005imputation}  & 20 minutes \\
\cite{troiano2008physical} & 60 minutes \\
\cite{robertson2010utility} & 20 minutes \\
\cite{evenson2011towards} & 20 minutes \\
\cite{schmid2015associations}  & 60 minutes\\
\cite{lee2016missing} & 20 minutes \\
\hline      
\end{tabular}
\caption{The definition of missing interval in terms of consecutive minutes of zeros in the literature on human activity.}
\label{missingness} %label after caption!!!!!
\end{table}

Table \ref{missingness} summarizes the related work on the length of an extended period of zero activity counts to be defined as missingness. In this paper, we choose to define missingness as a sustained interval of greater than or equal to 20 consecutive zero activity counts, which is the most commonly used criterion in the literature. Most missingness occurs between 10 pm to 8 am, which is the sleep time for most of the subjects. There is still sporadic missingness during other periods of time in the day, which may correspond to activities like swimming or bathing. The missingness periods are removed during the data preprocessing. The average proportion of zeros after removing the missingness is around 25\%, so that zero-inflation is still an issue to be considered in the modeling. In the data preprocessing, activity counts greater than 1,500 ($<$ 5\%) are truncated at 1,500 to ensure the numeric stability of the fitting algorithm.

\begin{table}[H]
\small
\centering
\begin{tabular}{lc}
\hline
Model specifications & BIC\\ \hline
5 states, type I & 248,009,082  \\
6 states, type I & 202,804,081  \\
7 states, type I & 203,261,150  \\
6 states, type II & 200,457,217  \\
6 states, type III &198,808,738  \\
6 states, type IV & 198,807,080  \\
\hline
\end{tabular}
\caption{Summary of BIC from model selection. In type I models, all parameters are subject-specific. In type II models, all parameters are subject-specific except the slopes, which are population parameters. In type III models, the intercepts are subject-specific; the slopes are population; the initial probabilities and transitions are subgroup-specific. In type IV models, all parameters are subgroup-specific except the intercepts, which are subject-specific. }
\label{bic}
\end{table}

\begin{figure}[H]
\centering
\includegraphics[scale=0.35]{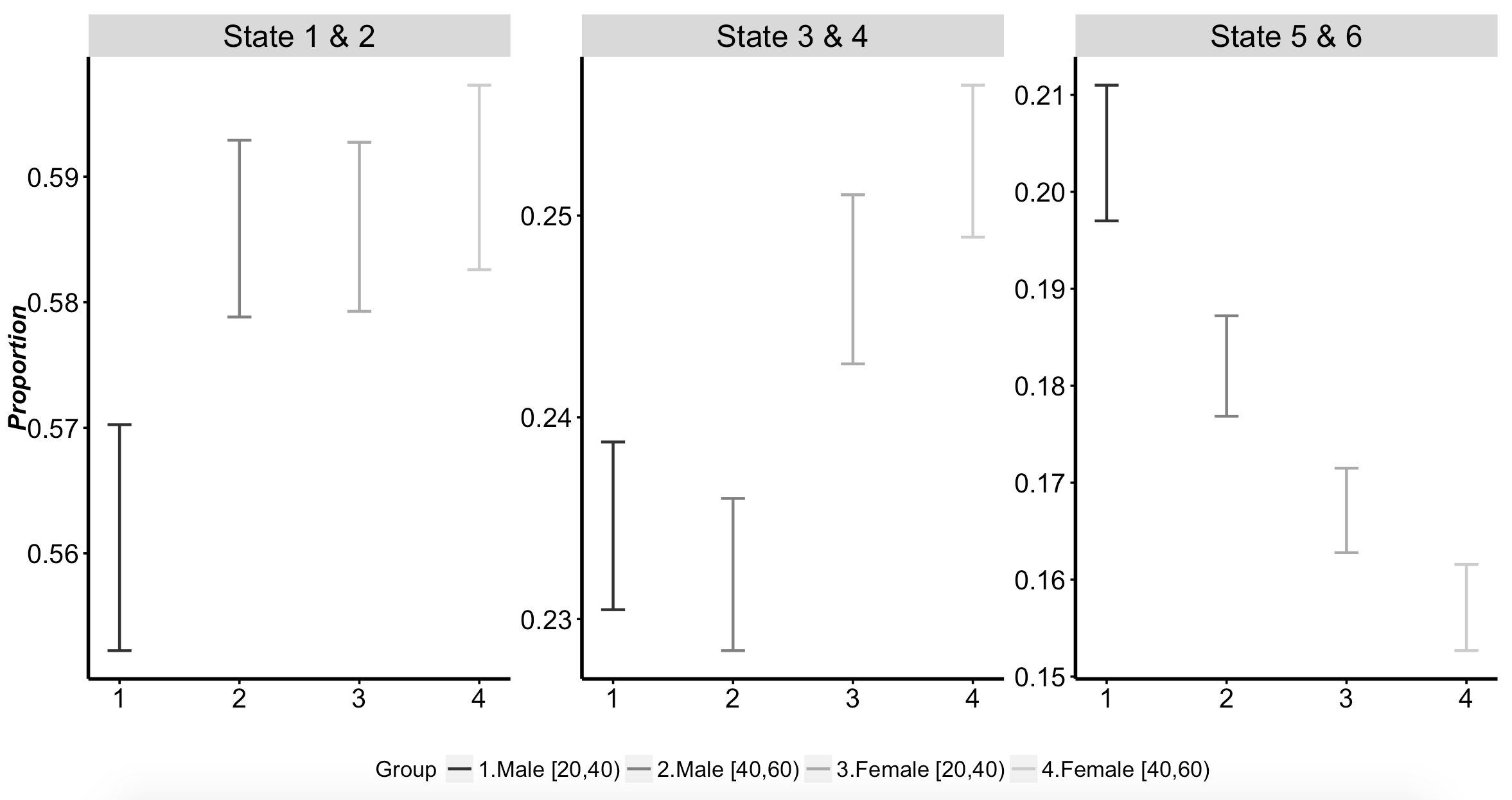}
\caption{ The 99\% bootstrap confidence intervals for the estimated proportion of time spent in latent activity states by subgroup in 03 - 04 NHANES.}
\label{resultplot}
\end{figure}

\begin{figure}[H]
\centering
\includegraphics[scale=0.35]{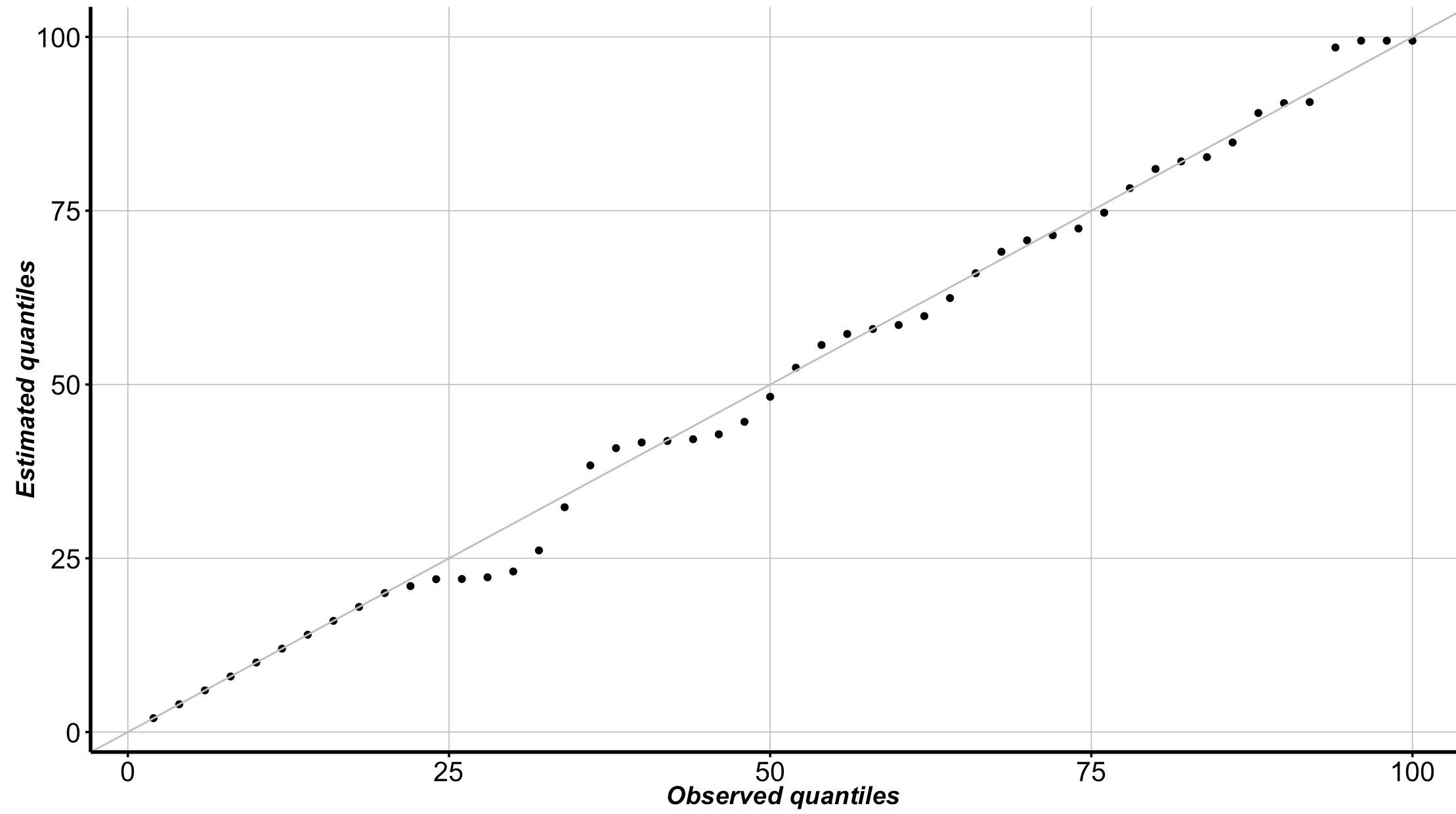}
\caption{ The estimated quantiles versus the observed quantiles of the accelerometer data in 03 - 04 NHANES, which shows no evidence of lack of fit. }
\label{qqplot}
\end{figure}

\begin{figure}[H]
\centering
\includegraphics[scale=0.35]{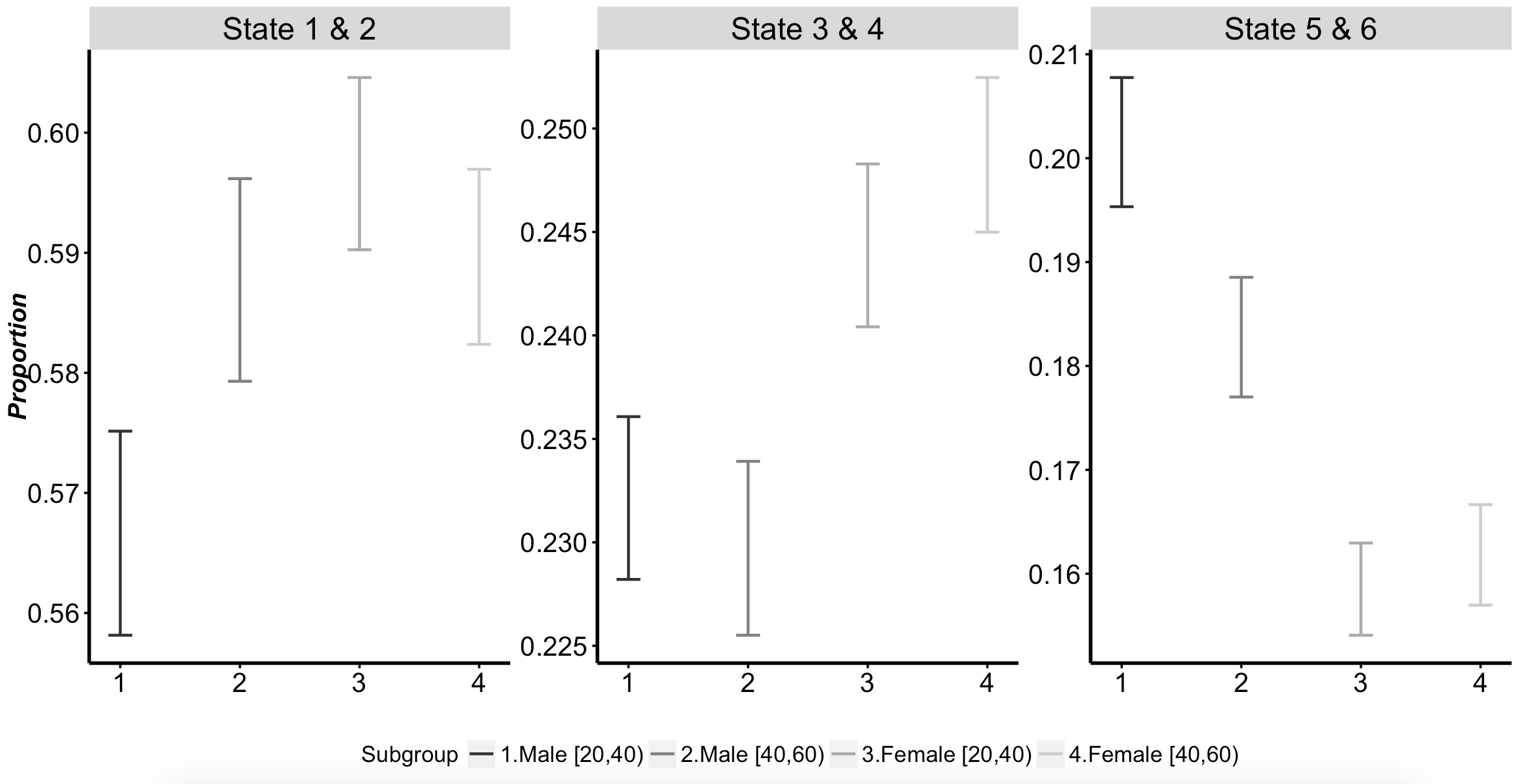}
\caption{ The 99\% bootstrap confidence intervals for the estimated proportion of time spent in latent activity states for each subgroup in 05 - 06 NHANES.}
\label{resultplot0506}
\end{figure}

To apply the HCTHMM model on the activity counts data, we need to select the number of latent states as well as the hierarchy for different sets of the parameters. The weekend effect is adjusted for in the Poisson and zero-inflated Poisson regression on the activity counts in each latent state. By the minimum BIC criterion as shown in Table \ref{bic}, we select the type IV HCTHMM with 6 latent states, where the intercepts in the state-dependent generalized linear models for logit zero proportion in state 1 and log Poisson means in the all states are subject-specific, while the initial probabilities, transition rates, and the slopes in the state-dependent generalized linear models are subgroup-specific. This final model indicates the baseline zero proportion and mean activity counts in each latent state vary across subjects. For all the other parameters, the between-subgroup variability is more prominent than the within-subgroup variability.
Figure \ref{resultplot} shows the 99\% confidence interval for the estimated proportion of time spent in latent activity states for each subgroup in 03 - 04 NHANES. There are several interesting findings. First, younger men spend less time in the low intensity activity states (state 1, 2) than older men and women. Second, men spend less time than women in the medium intensity activity states (state 3, 4). Third, men spend more time than women in the high intensity activity states (state 5, 6). 
Figure \ref{qqplot} plots the estimated quantiles versus the observed quantiles for the accelerometer data,
which aligns well along the 45-degree line, indicating no lack of fit.
To validate the results, we apply the HCTHMM methodology to 05 - 06 NHANES, which has the same study setup and data structure as the 03 - 04 NHANES. 
Figure \ref{resultplot0506} shows the 99\% confidence interval for the estimated proportion of time spent in latent activity states for each subgroup in 05 - 06 NHANES,  which has a similar pattern as seen in 03 - 04 NHANES.

\section{Conclusions}

We propose HCTHMM to be valid inference strategy for the longitudinal activity data. Within this framework, we can estimate the mean state probabilities for different subgroups of subjects as well as quantify the uncertainty. Our findings are consistent with previous literature on human physical activity \citep[][]{metzger2008patterns,troiano2008physical,hansen2012accelerometer,xiao2014quantifying}, which indicated that the physical activity can be classified into different categories by intensity, and that the activity level decreases as a result of aging. Moreover, women tend to spend more time in lighter intensity activity, whereas younger men tend to have periods of higher intensity activities.
 
In the future, this HCTHMM framework can be extended to the controlled clinical studies to estimate certain treatment effects in a specific cohort of patients. We can also allow for time-varying covariates in the transition rates. Moreover, when some model parameters are truly subject-specific or subgroup-specific, it may be more powerful to model them as random so that tests based on variance components can be constructed to test their effects. Another modification is to extend the latent continuous-time Markov process to a semi-Markov process. This will be scientifically interesting because it is reasonable to assume that the current latent state not only depends on the most recent past state but also on the history of the state trajectory. However, all these changes are computationally expensive, especially on such large-scale high-frequency data. Corresponding estimation methods have to be developed before the application becomes feasible.

\bibliographystyle{Chicago}
\bibliography{bibliograph}

\end{document}